# Detection of hydrogen by the extraordinary Hall effect in CoPd alloys.

S. S. Das, G. Kopnov and A. Gerber

Raymond and Beverly Sackler Faculty of Exact Sciences,

School of Physics and Astronomy, Tel Aviv University,

Ramat Aviv, 69978 Tel Aviv, Israel

Effect of hydrogen adsorption on the extraordinary Hall phenomenon (EHE) in ferromagnetic CoPd films is studied as a function of composition, thickness, substrate and hydrogen concentration in atmosphere. Adsorption of hydrogen adds a positive term in the extraordinary Hall effect coefficient and modifies the perpendicular magnetic anisotropy with the respective changes in coercivity and remanence of hysteresis loops. Hydrogen sensitive compositions are within the Co concentration range 20% $\leq x \leq$ 50% with the strongest response near the EHE polarity reversal point $x_0 \sim 38\%$. Depending on the film composition and field of operation the EHE response of CoPd to low concentration hydrogen can reach hundreds percent, which makes the method and the material attractive for hydrogen sensing.

## Introduction.

Hydrogen is a combustible gas present in practically every chemical process. The detection and concentration measurement of hydrogen is of a paramount importance in limitless cases of human activity from chemical, metallurgical, semiconductor and nuclear power industry to the emerging hydrogen energy economy. Many types of hydrogen sensors are commercially available or are in development. Following the classification by Hübert et al[1] they can be divided in eight groups as: catalytic, thermal conductivity; electrochemical, resistance based, work function based, mechanical, optical and acoustic. Yet, there is a continued need for faster, more accurate and more selective detection of hydrogen gas. It was suggested recently[2] that accuracy and selectivity of gas detection in general and hydrogen in particular, could be improved by measuring two or more independent gas-dependent parameters, e.g. resistance and magnetization. To execute a sensitive magnetic measurement in a compact and handy apparatus it was proposed to use the extraordinary Hall effect (EHE), which is an electric replica of magnetization compatible with a standard four-probe resistance measurement. Successful implementation of the technique would enrich the gas detection arsenal by magnetic type of sensors using the spintronics effect. In this paper, we present a systematic study of hydrogen detection using the extraordinary Hall effect in thin CoPd films.

## Experimental.

Polycrystalline $Co_xPd_{(100-x)}$ films with Co atomic concentration $x$ (at. %) varying over an entire range $0 \le x \le 100$ were deposited by rf co-sputtering from Co and Pd targets onto room temperature glass and GaAs substrates. The base pressure prior to deposition was $5 \times 10^{-7}$ mbar. Sputtering was carried out at Ar-pressure of $5 \times 10^{-3}$ mbar. Composition of samples was controlled by rf-power of the respective sputtering sources. Co and Pd are soluble and form an equilibrium fcc solid solution phase at all compositions[3] during the room temperature deposition. Four series of samples with nominal thicknesses of 7 nm, 14 nm, 70 nm, and 100 nm were prepared. Resistivity and Hall effect measurements were done at room temperature by using Van der Pauw protocol. The set up was equipped with a gas-control chamber, which enabled performing measurements at variable hydrogen concentrations up to 4%. Magnetization of 100 nm thick CoPd samples was measured in a SQUID magnetometer.

## Results and discussion.

1. Magnetization and EHE in CoPd alloys.

The field dependent Hall resistivity in ferromagnetic films can be presented as:

$$\rho_H(B) = R_{OHE}B + \mu_0 R_{EHE} M(B) \qquad (1)$$

where *B*, and *M* are components of the magnetic induction and magnetization normal to the film plane, $R_{OHE}$ is the ordinary Hall effect coefficient related to the Lorentz force acting on moving charge carriers and $R_{EHE}$ is the extraordinary Hall effect coefficient associated with a break of the time reversal symmetry at spin-orbit scattering in magnetic materials. The EHE contribution can exceed significantly the ordinary Hall effect term in the relevant low field range, and the total Hall resistance $R_H$ can be approximated as:

$R_H = \rho_H/t = \mu_0 R_{EHE} M/t$, where $t$ is the film thickness. Coefficient $R_{EHE}$ is assumed to be field independent, therefore the field dependence of the Hall signal is directly proportional to the normal to plane magnetization. Magnitude of the signal depends on magnetization and $R_{EHE}$. CoPd is one a few ferromagnetic materials, in which the coefficient $R_{EHE}$ varies strongly with composition and even reverses its polarity [4]. Fig.1 illustrates this unusual property and its implications. Magnetization of three 100 nm thick $Co_xPd_{100-x}$ films with Co atomic concentration x = 25%, 36% and 47% is shown in Fig.1a as a function of field applied normal to the film plane. Magnitude of the saturated magnetization increases linearly with Co concentration. Perpendicular magnetic anisotropy and the respective hysteresis in the field dependent magnetization were attributed [4,5] to an interfacial strain in CoPd, which is known to have a very large magnetostriction [6,7]. Fig.1b presents the EHE resistivity as a function of magnetic field measured in the same samples. Contrary to magnetization, the EHE reverses its polarity $d\rho_H/dB$ from positive in the x = 47% sample to negative in films with x = 25% and 36%. Despite a change of polarity, the EHE signal remains an electrical replica of the corresponding magnetization loop in all samples, as demonstrated in Fig.1c. Here, the normalized magnetization $M/M_{sat}$ and the normalized EHE resistivity $\rho_H/\rho_{H,sat}$, where $M_{sat}$ and $\rho_{H,sat}$ are the saturated magnetization and EHE resistivity respectively, are plotted for two samples: x = 36% with a negative EHE and x = 47% with a positive EHE. The normalized magnetization hysteresis loops are identical to the respective EHE ones for all samples. One can, therefore use the EHE measurements to determine magnetic parameters of the material: the field dependence of magnetization, coercive field and squareness of the hysteresis loops.

The saturated EHE resistivity $\rho_{H,sat}$ of three series of samples with thickness $t$ = 7 nm, 14 nm and 70 nm is plotted in Fig.2 as a function of Co concentration. $\rho_{H,sat}$ is determined by extrapolating the linear high field portion of $\rho_H(B)$ curves to zero field. The $\rho_{H,sat}$ behavior is similar in all series: EHE is absent in pure Pd films at x = 0%; negative EHE is developed in Pd-rich samples reaching $\rho_{H,sat} \approx$ - 0.1 μΩcm at x ~ 25%; $\rho_{H,sat}$ varies linearly with $x$ at higher Co concentrations up to + 0.3 μΩcm at x ~ 70%. The slope $d\rho_{H,sat}/dx$ is about 0.01 μΩcm/Co%. Polarity of the signal reverses to positive at Co concentration of about 38%, the concentration we denote as $x_0$.

Reversal of the EHE polarity with composition has been reported in Co-Pd systems such as CoPd alloys,[4,8] Co/Pd multilayers[9-14] and also in other ferromagnetic alloys: NiFe,[15] TbCo[16] *etc.*. The split band model with two partially filled *3d* bands was used [15] to explain the effect in NiFe. For material like MnGaAs with an impurity band, the EHE conductivity is expected[17] to be proportional to the derivative of the density of states at the Fermi energy and, therefore to change sign as the Fermi level crosses the density-of-states maximum. Adaptation of these models to CoPd is questionable since the polarity reversal was also observed with aging[4,10] as a function of temperature[13], the repetition number and a relative Co and Pd layers thickness in multilayers[9,13]. Thus, comprehensive understanding of the EHE polarity reversal in these materials is lacking.

Characteristics of the EHE hysteresis loops as a function of Co concentration are collected in Fig.3. Fig.3a presents the remnant EHE resistivity $\rho_{H,rem}$ measured at zero magnetic field after sweeping the field from the positive 1.5 T value down to zero. $\rho_{H,rem} = 0$ in samples without hysteresis, $\rho_{H,rem} < 0$ in films with a negative EHE polarity, and $\rho_{H,rem} > 0$ in films with a positive EHE polarity. Magnitudes of the remnant EHE resistivity $\rho_{H,rem}$ (Fig.3a) and the saturated resistivity $\rho_{H,sat}$ (Fig.2) are close in series of all thicknesses. Figs. 3b presents the coercive field $B_c$. Hysteresis is found in the concentration range 20% ≤ $x$ ≤ 50%. Thicker films show higher coercivity up to about 100 mT in the 70 nm thick series, which is about three times larger than the highest observed in the 7 nm thick one. Squareness $S$ of hysteresis loops, which characterizes the degree of perpendicular anisotropy, was defined by the ratio: $S = {\rho_{H,rem}}/{\rho_{H,sat}}$. $S$ reaches unity in 7 nm and 14 nm thick series.

As mentioned above, it was suggested that the source of perpendicular anisotropy in CoPd films is an interfacial strain in films with strong magnetostriction. In this case, the substrate can affect the resulting anisotropy and hysteresis. Indeed, films deposited on crystalline GaAs substrates demonstrate higher perpendicular anisotropy than those deposited on amorphous glass substrates. The difference is illustrated in Fig.4 presenting squareness of 7 nm films deposited on glass and GaAs as a function of Co content. Films deposited on GaAs show a full remanence in an almost entire concentration range where hysteresis is observed. Films grown on glass exhibit hysteresis with reduced remanence for 30% < x < 50%. Otherwise, properties of films grown on glass and GaAs are similar, and we will not emphasize the substrate in the following.

2. Hydrogenation effect on EHE in CoPd alloys.

To establish an effect of hydrogen on CoPd alloys, we performed the EHE measurements in ambient air, vacuum, nitrogen and in hydrogen-nitrogen mixture with various hydrogen concentrations up to 4%. The field dependent EHE loops were identical when measured in vacuum, air and nitrogen atmosphere. However, the response is significant in the $H_2/N_2$ mixture gas. Figures 5 (a) – (d) present the EHE loops of four selected 7 nm thick $Co_xPd_{(100-x)}$ samples with x = 28%, 37%, 41% and 47% in ambient air and in 4 % $H_2/N_2$ mixture gas at room temperature. Exposure to hydrogen modifies the magnitude of the EHE signal, the form and width of the hysteresis and, strikingly, the very polarity of the EHE in the 37% sample. When exposed to hydrogen, the absolute value of $\rho_{H,sat}$ decreases in the sample x = 28% which exhibits a negative EHE in air; increases in x = 41% and 47%, that have a positive EHE in air; and the signal reverses from negative to positive in x = 37%. Thus, in all samples hydrogenation induces a positive shift (at positive fields) in the EHE resistivity $\Delta\rho_{H,sat} = \rho_{H,sat}(H_2) - \rho_{H,sat}(air)$. The effect can be understood following the model of Ref. [18], that assumes that Matthiessen's rule can be applied not only to resistivity but also to EHE. Different spin-orbit scattering sources contribute independently to the EHE, and the final signal is a sum of all contributions. In the present case, the hydrogen induced EHE term is positive and independent on polarity of the signal in air, the latter being positive or negative depending on the alloy composition. Positive polarity of $\Delta\rho_{H,sat}$ is consistent with a change in an effective Co/Pd ratio upon hydrogenation of an alloy. Assuming that Pd bonds to hydrogen, the effective Co vs Pd concentration increases by $\Delta x_{eff}$, that gives a positive change in the saturated EHE resistivity: $\Delta\rho_{H,sat} = {d\rho_{H,sat}}/{dx}\ \Delta x_{eff}$. For ${d\rho_{H,sat}}/{dx}$ = 0.01 μΩcm/Co% (Fig.2), one estimates $\Delta x_{eff}$ ~ 1 %.

The absolute $\Delta\rho_{H,sat}$ and the normalized ${\Delta\rho_{H,sat}}/{\rho_{H,sat}(air)}$ response of the saturated EHE resistivity to hydrogen are shown in Fig.6 as a function of composition for series of different thickness. Significant hydrogen effect is limited to Co concentration range 20% ≤ x < 60%. The thinnest 7 nm films demonstrate the largest absolute response to hydrogen with a clear maximum in the vicinity of the polarity reversal concentration $x_0$, the latter is common to series of all thicknesses. The highest normalized response of the saturated EHE resistivity in this series is about 130% in $Co_{0.37}Pd_{0.63}$ sample with the lowest saturated EHE resistivity $\rho_{H,sat}(air)$. Notably, by fine-tuning the CoPd composition to the reversal concentration $x_0$, one can obtain zero EHE resistivity in air, in which case the ratio $\Delta\rho_{H,sat}/\rho_{H,sat}(air)$ is unlimited.

Hydrogenation has a strong impact on magnetic anisotropy and, thus, on the form of the EHE hysteresis loops. Examples of these changes can be seen in Fig.5 (a, b and c). The coercive field, the squareness and the ratio between them are affected by hydrogen. In most samples hysteresis loops shrink in hydrogen and coercivity decreases compared to its value in air. However, shrinking of hysteresis with hydrogenation is not a general property. Fig.4c presents the EHE loops of 7 nm thick sample with x = 41%. Here, all parameters of the hysteresis: the coercive field, remanence, squareness and the hysteresis closure field in hydrogen are larger than in air. Similar properties are also exhibited by the sample grown on GaAs substrate. Recent Kerr effect study of CoPd alloys hydrogenation [19] found an increase of coercivity in hydrogen. Thus, further study of the phenomenon is needed.

The normalized changes of coercive field ($\Delta B_c/B_c(air)$), and the remnant EHE resistivity ( $\Delta\rho_{rem}/\rho_{rem}(air)$) are summarized as a function of composition in Fig.7. Expansion of hysteresis loop in hydrogen in the x = 41% sample is seen as an effect of an opposite polarity compared with rest of data. The largest absolute changes of coercive field are found in the 14 nm thick series. The largest relative changes are in the 7 nm thick series in samples x = 24% (negative) and x = 41% (positive) that are at the edges of the hysteresis range where the derivative $dB_c/dx$ are the largest (see Fig.3b). The remanence response to hydrogen (Fig.7b) is similar in series of all thicknesses, excluding the expanding x = 41% 7 nm thick sample.

Sensitivity to hydrogen concentration was tested by performing the field dependent EHE measurements in atmosphere containing different amounts of hydrogen in the range 0 – 4 %. At first, the chamber was filled with nitrogen and the measurement was done in $N_2$ atmosphere. For the subsequent measurement at the desired hydrogen concentration the required fraction of nitrogen was replaced by the 4 % $H_2/N_2$ mixture. At the end of each measurement at a particular $H_2$ concentration, the sample was exposed to ambient air for dehydrogenation. After completing the EHE measurements at different $H_2$ concentrations, the sample was measured again in $N_2$ atmosphere, and the data were fully reproducible. Figure 8 shows typical EHE loops measured at hydrogen concentrations y = 0%, 0.2%, 1% and 4% in the 7 nm thick x = 32 % sample grown on a GaAs substrate. The EHE resistivity data at three fixed fields B = 0T, -13 mT and 0.1 T are shown as a function of hydrogen concentration $y$ in Fig. 9. B = 0T data correspond to the remnant EHE resistivity $\rho_{H,rem}$, B = 0.1T data is that of the saturated EHE resistivity $\rho_{H,sat}$ and B = -13 mT data correspond to the field within the hysteresis loop with the largest EHE resistivity change $\Delta\rho_H = \rho_H(H_2 4\%) - \rho_H(0)$, where $\rho_H(0)$ is the EHE signal in air at a given field. The results are presented in the form of the normalized hydrogen induced changes, as: $\Delta\rho_{H,norm} = \frac{\Delta\rho_H(y)}{\rho_H(0)} = \frac{\rho_H(y)-\rho_H(0)}{\rho_H(0)}$. $\Delta\rho_{H,norm}$ of this sample at 4% $H_2/N_2$ mixture is 16% in the saturated state at 0.1 T, 22% in the remnant state at zero field and 182% under a fixed field of -13 mT. Thus, the changes are the largest within the hysteresis loop, caused by reduction of the coercive field and an effective reversal of magnetization. The highest sensitivity of all measured parameters to hydrogen is at the lowest hydrogen content, already below 0.2%. Sensitivity to hydrogen defined as $d\Delta\rho_{H,norm}/dy$ below 0.2% is higher than 500 %/$10^4$ ppm at B = - 13 mT (dashed line in Fig.9a) .

Analytical presentation of the results is ambivalent due to a limited range of data and hydrogen concentrations (about a decade). At hydrogen concentrations 0.2% ≤ y ≤ 4% $\Delta\rho_{H,norm}$ can be well fitted by an exponent: $\Delta\rho_{H,norm} = A + Bexp\left(-\frac{y}{Y}\right)$ with A = -182 (B=-13 mT), -22 (B = 0mT) and -16 (B = 0.1 T) (solid curves in Fig.9a). The same data can be alternatively presented by the power law: $\Delta\rho_{H,norm} = \alpha y^n$ with $n$ = 0.2 and $\alpha$ = 142 (B=-13mT), 16 (B=0 mT) and 14 (B= 0.1 T) (Fig.9b). The same uncertainty holds in presentation of the coercive field as a function of hydrogen concentration. $B_c$ can be well fitted by an exponent: $B_c(y) = C + Dexp(y/Y)$, $(C + D = B_C(0))$ with $Y$=2.0, or by the power law expression: $B_c(y) = {B_C}^*(0)(1 + y)^{(-n)}$ with ${B_C}^*{\sim}14\ mT$ and $n{\sim}0.2$.

An important property of a sensor is reversibility after cyclic replacement of clean air and an atmosphere containing hydrogen. For simplicity of operation, an EHE-based gas sensor is expected to work at a fixed magnetic field either within the hysteresis range or under field high enough to saturate the magnetization. Fig.10 presents a typical EHE response to a consequent exposure of a sample (x=28%) to 4 % $H_2/N_2$ mixture followed by refilling the chamber with an ambient air at three fixed fields: $B_1$ = 0, $B_2$ = 10 mT and $B_3$ = 0.5 T. $B_1$ and $B_2$ are within the hysteresis loop and $B_3$ is beyond the hysteresis in the saturated state. The sample responds to hydrogen at all three fixed fields and the changes correspond to the values obtained in the field dependent hysteresis loop in hydrogen atmosphere. When hydrogen atmosphere is removed and replaced by air, $\rho_{H,sat}$ at $B_3$ recovers to its original value at air, while the signals measured at $B_1$ and $B_2$ within the hysteresis loop do not recover. Exposure to hydrogen at low external fields leads to an effective demagnetization of the material. Refilling by air doesn't recover the ordered magnetic state, and only minor changes are observed in the remnant signal $\rho_{H,rem}$ due to change in the coefficient $R_{EHE}$. Magnetization and the respective EHE signal can be recovered to the original air values only by re-magnetizing the material in high field. This is demonstrated in Fig.11. The first hysteresis loop (open circles) was measured in air starting from a magnetically disordered state O. The loop ends in the remnant state at B = 0 marked by letter A. Next, the chamber was filled with the hydrogen mixture and the measured signal moved to the value marked C. Then, hydrogen was removed and replaced by air, while the EHE signal remained at C. The second hysteresis loop in air, marked by a solid blue line, started at C, reached saturation at field above 70 mT and recovered to its pre-hydrogenation state under further field sweeping.

## Summary.

CoPd alloy is an outstanding ferromagnetic system in which polarity of the extraordinary Hall effect changes from positive in Co-rich alloys to negative in Pd-rich ones at the polarity reversal concentration $x_0$. Adsorption of hydrogen causes modifications in both magnetic and Hall effect properties of the films. Hydrogen induced spin-orbit scattering adds a positive term to the extraordinary Hall effect coefficient $R_{EHE}$, which leads to a reduction of the EHE resistivity in films with negative $R_{EHE}$, increase of $\rho_{H,sat}$ where the coefficient is positive, and reversal of the signal from negative in air to positive in hydrogen in the vicinity of $x_0$. Hydrogen also affects the perpendicular anisotropy and the respective field dependent hysteresis by modifying the form, coercivity and remanence. Hydrogen sensitive compositions are within the Co concentration range $20\% \leq x \leq 50\%$ with the strongest response near the EHE polarity reversal point $x_0 \sim 38\%$. Depending on the film composition, thickness and field of operation the EHE response of CoPd to low concentration hydrogen can reach hundreds percent, which makes the method and the material attractive for hydrogen sensing.

The research was supported by the State of Israel Ministry of Science, Technology and Space grant No.53453.

## References.

## Figure Captions.

Fig.1. Field dependent magnetization (a); Hall resistivity (b); the normalized magnetization ${M}/{M_{sat}}$ and the normalized EHE resistivity ${\rho_H}/{\rho_{H,sat}}$ (c) of three 100 nm thick $Co_xPd_{100-x}$ films with Co atomic concentration x = 25%, 36% and 47%. Subscript "sat" indicates the respective saturated values at high field. Fig.1c presents the data for x = 36% and 47%. Field is applied normal to plane. Solid lines in (a) and (b) are guides for the eye.

Fig.2. The saturated EHE resistivity $\rho_{H,sat}$ of three series of samples with thickness $t$ = 7 nm, 14 nm and 70 nm as a function of Co atomic concentration $x$.

Fig.3. The remnant EHE resistivity $\rho_{H,rem}$ (a); and coercive field $B_c$ (b) in series of 7nm, 14 nm and 70 nm thick $Co_xPd_{100-x}$ films as a function of Co concentration $x$. Dashed lines are guides for the eye.

Fig.4. Squareness of 7 nm thick $Co_xPd_{100-x}$ films deposited on glass (open triangles) and GaAs (solid circles) substrates as a function of Co concentration x.

Fig.5. EHE resistivity as a function of normal to plane magnetic field in four 7 nm thick $Co_xPd_{100-x}$ samples with x = 28% (a), 37% (b), 41% (c) and 47% (d) measured in air (open circles) and in hydrogen/nitrogen mixture with 4% of $H_2$.

Fig.6. The absolute (a) and the normalized (b) changes of the saturated EHE resistivity in 4% $H_2/N_2$ atmosphere in 7nm, 14 nm and 70 nm thick $Co_xPd_{100-x}$ films as a function of Co concentration $x$.

Fig.7. The normalized changes of coercive field (a) and the remnant EHE resistivity at zero magnetic field (b) in 4% $H_2/N_2$ atmosphere in 7nm, 14 nm and 70 nm thick $Co_xPd_{100-x}$ films as a function of Co concentration *x*.

Fig.8. EHE resistivity as a function of magnetic field in 7 nm thick $Co_{32}Pd_{68}$ sample on GaAs substrate measured in air (y = 0) and in hydrogen/nitrogen atmosphere with different hydrogen concentrations between 0.2% up to 4%. The vertical solid line corresponds to field B=-13 mT where the largest fixed field change in EHE resistivity is observed.

Fig.9. The normalized hydrogen induced changes in the EHE resistivity $\Delta\rho_{H,norm}$ as a function of hydrogen concentration y for $Co_{32}Pd_{68}$ sample at fixed fields B = 0T (open stars) and -13 mT (open circles) within the hysteresis loop and B = 0.1 T (open triangles) in the saturated range. (a) Solid curves between y = 0.2% and 4% are fits to the exponential function $\Delta\rho_{H,norm} = A + Bexp\left(-\frac{y}{Y}\right)$. (b) The same data presented in log-log scale with straight lines corresponding to the power law fit: $\Delta\rho_{H,norm} = \alpha y^n$ with $n\sim0.2$.

Fig.10 Normalized EHE resistivity of 7 nm thick $Co_{28}Pd_{72}$ sample during the sequential filling the chamber with 4% $H_2/N_2$ mixture and replacement the mixture by air under fixed fields: $B_1$=0T (a), $B_2$=10 mT (b) and $B_3$=0.5T (c).

Fig.11. EHE resistivity hysteresis loops measured in air before (open circles) and after (blue solid line) filling the chamber with 4% $H_2/N_2$ mixture and subsequent replacement the mixture by air. The first loop starts at point O and ends at A. The second loop starts at point C. Red dotted curve is the loop measured in 4% $H_2/N_2$ mixture.

**Figures.**

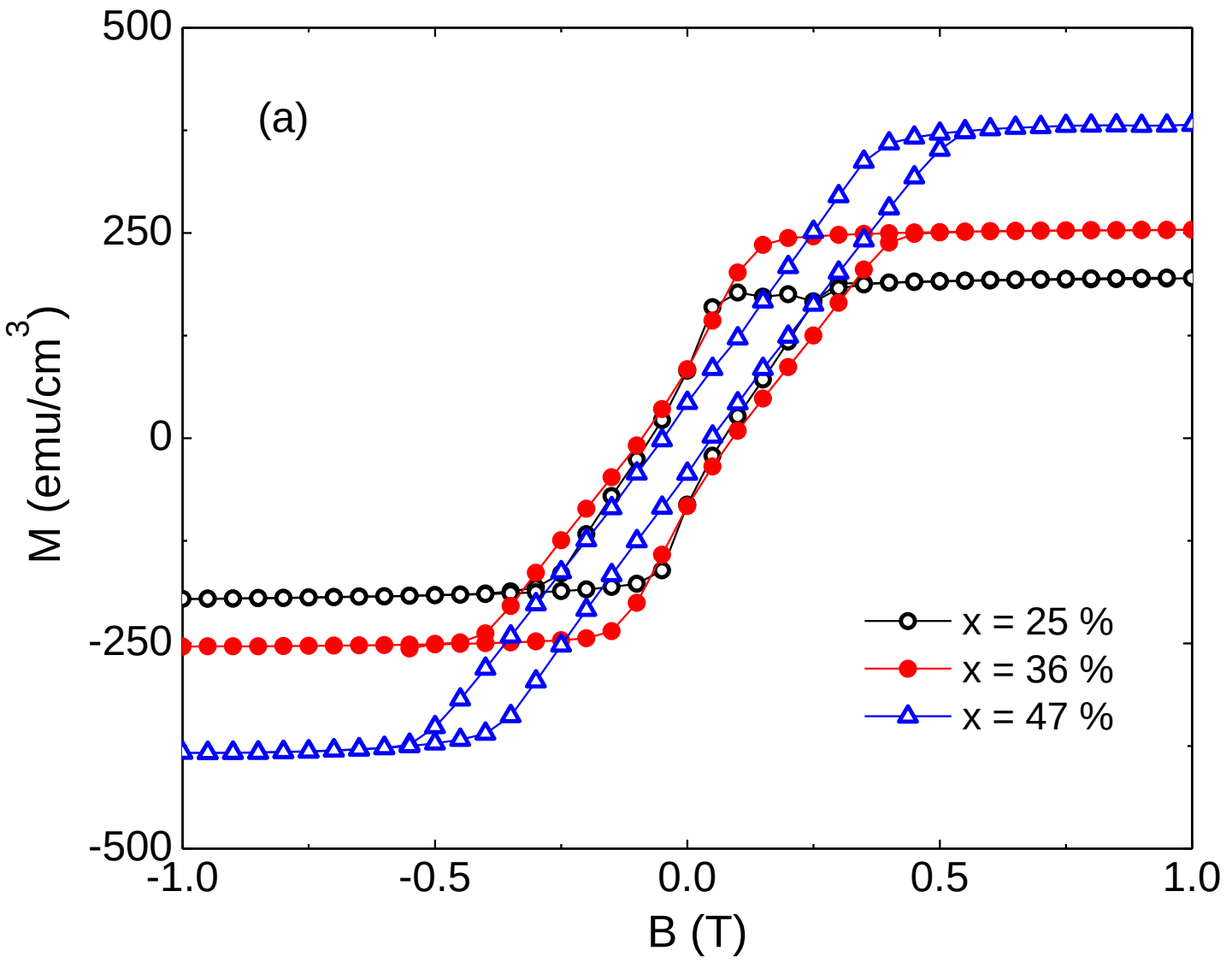

Fig. 1a

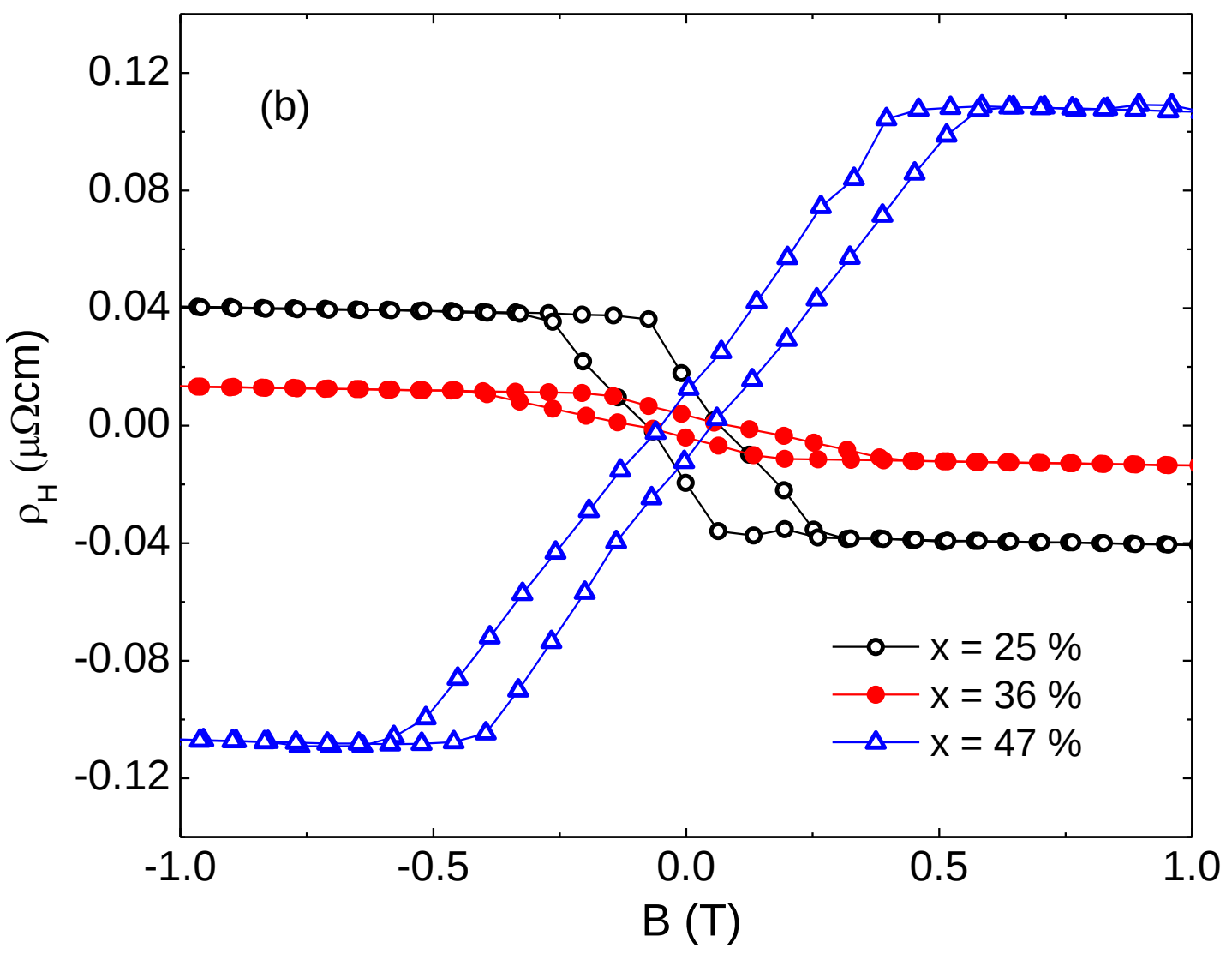

Fig. 1b

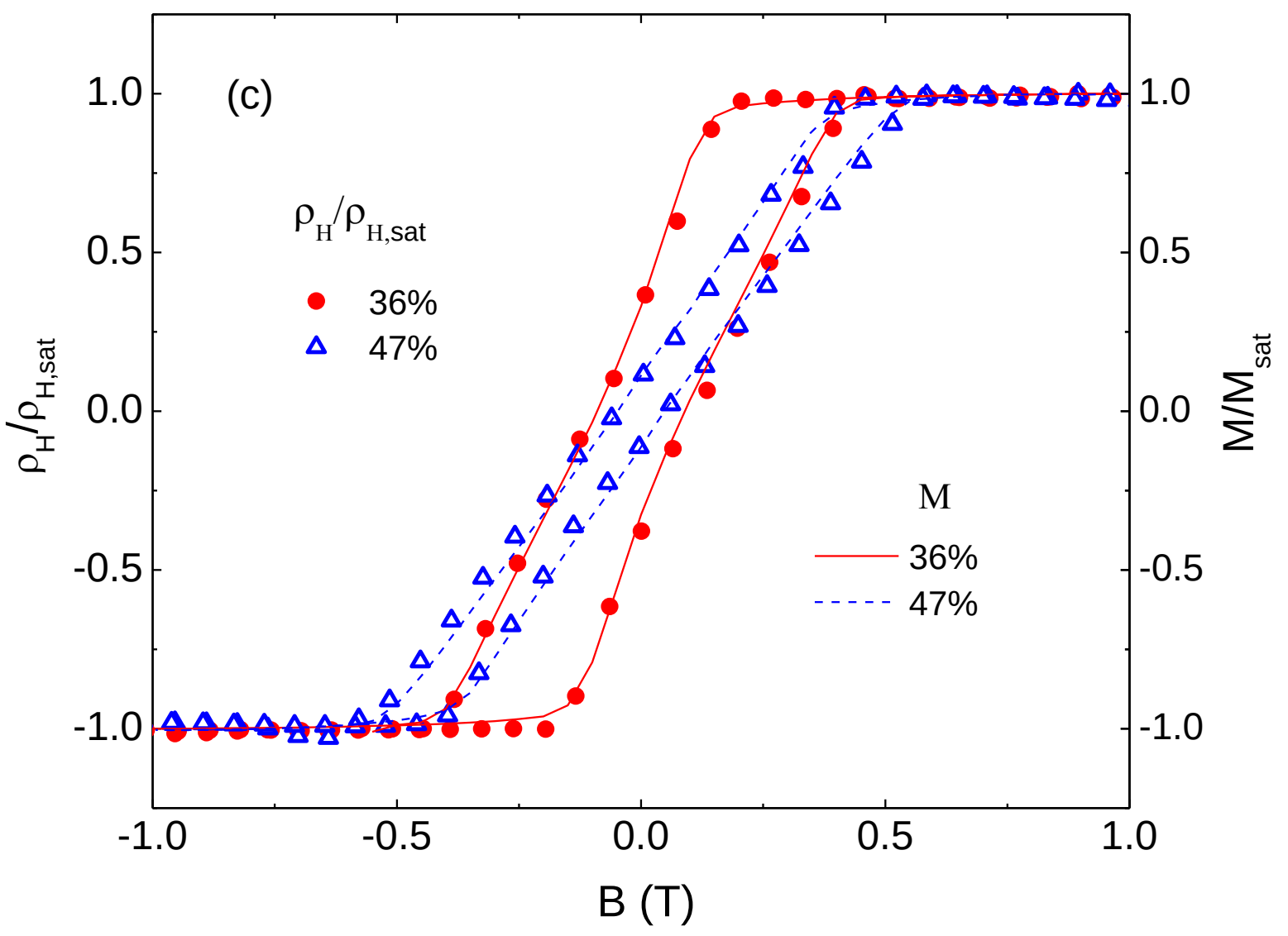

Fig.1c

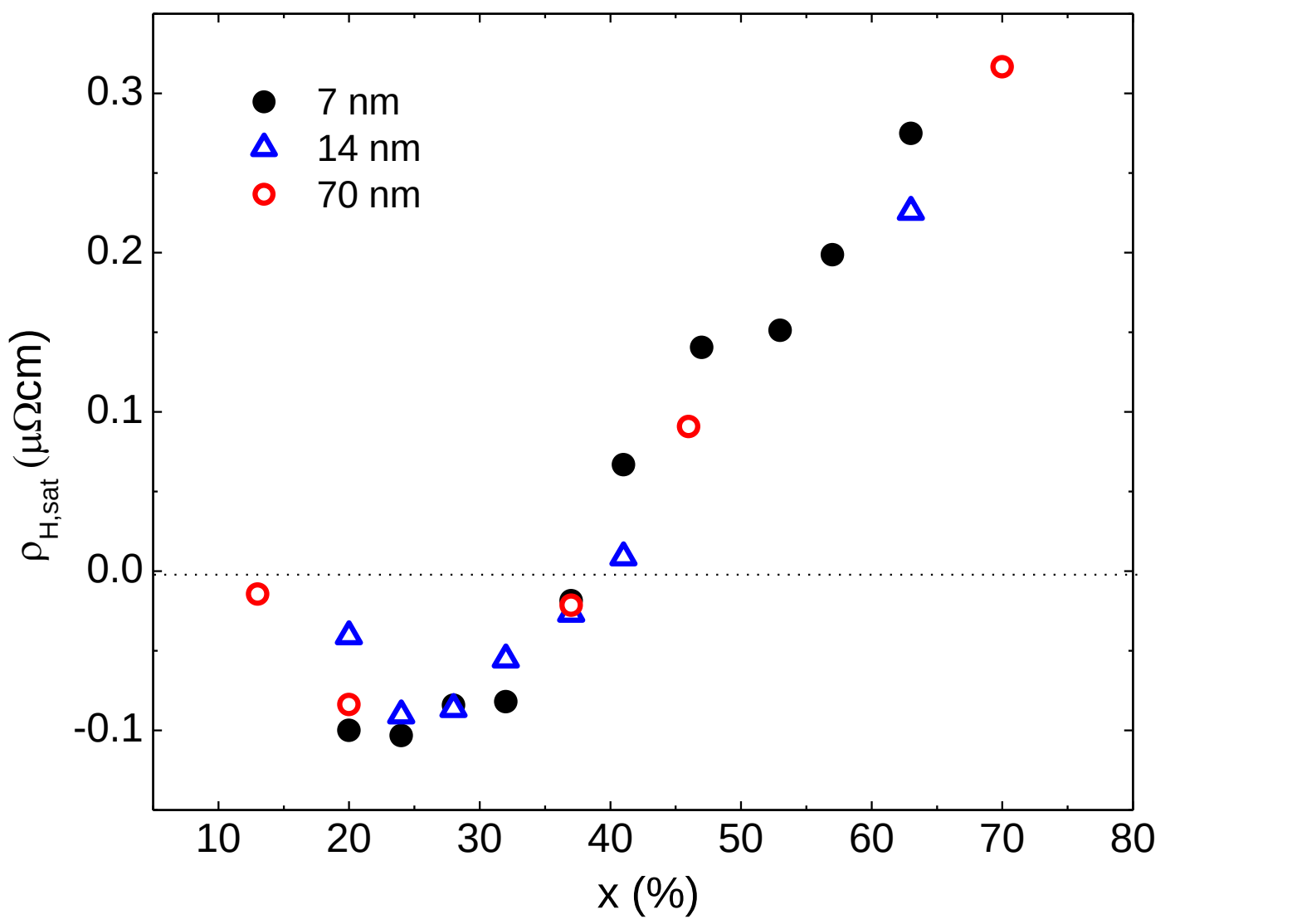

Fig.2.

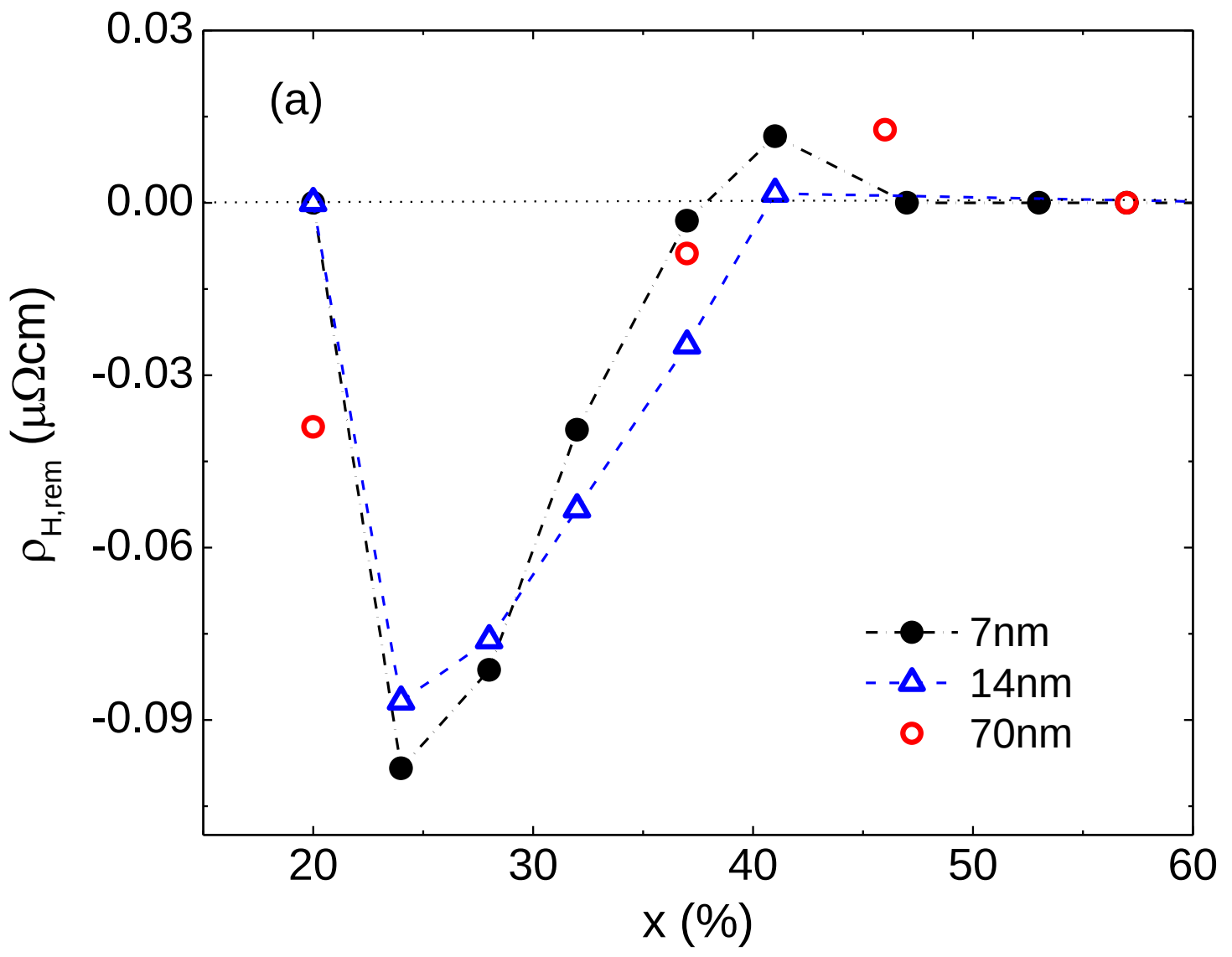

Fig. 3a

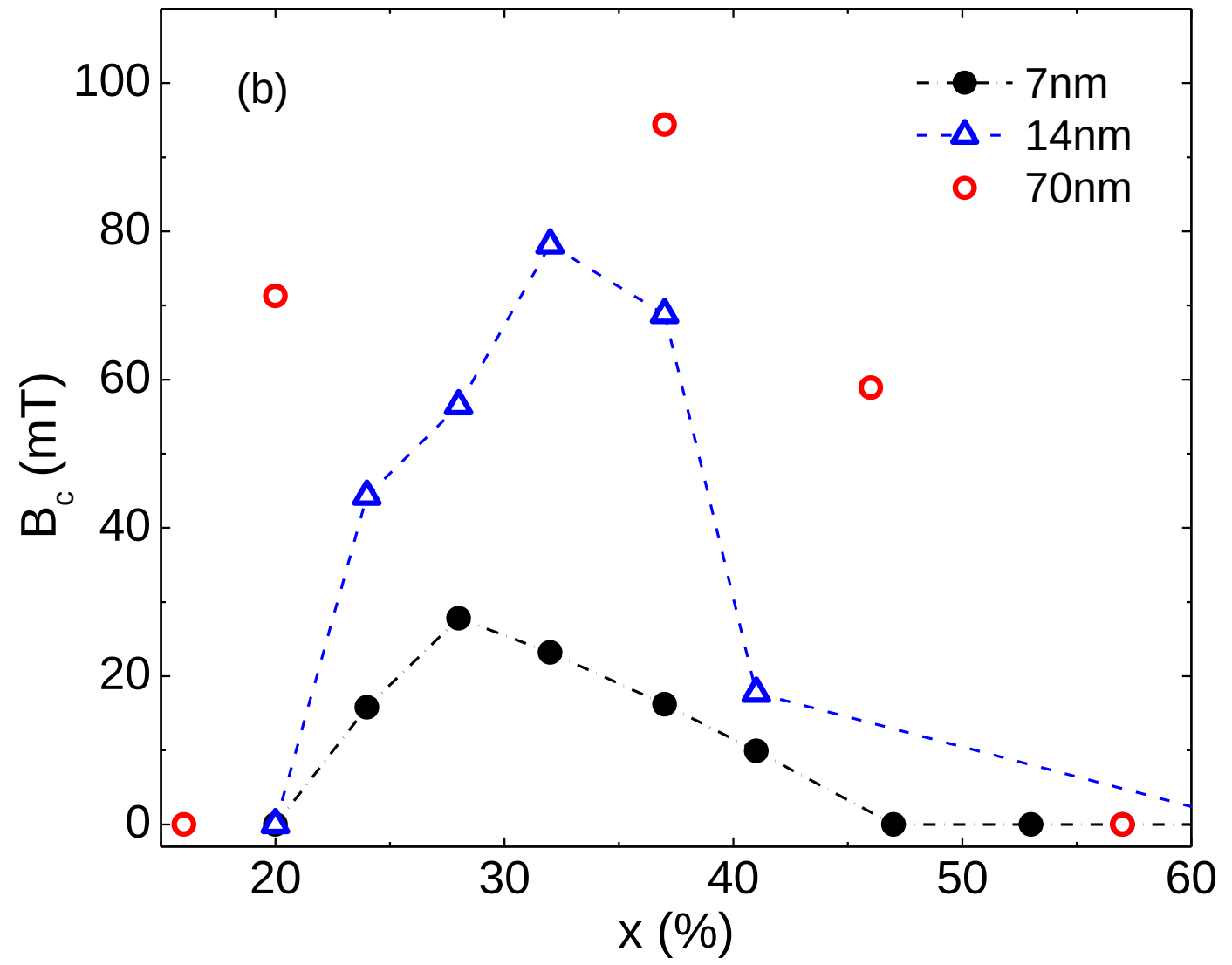

Fig. 3b

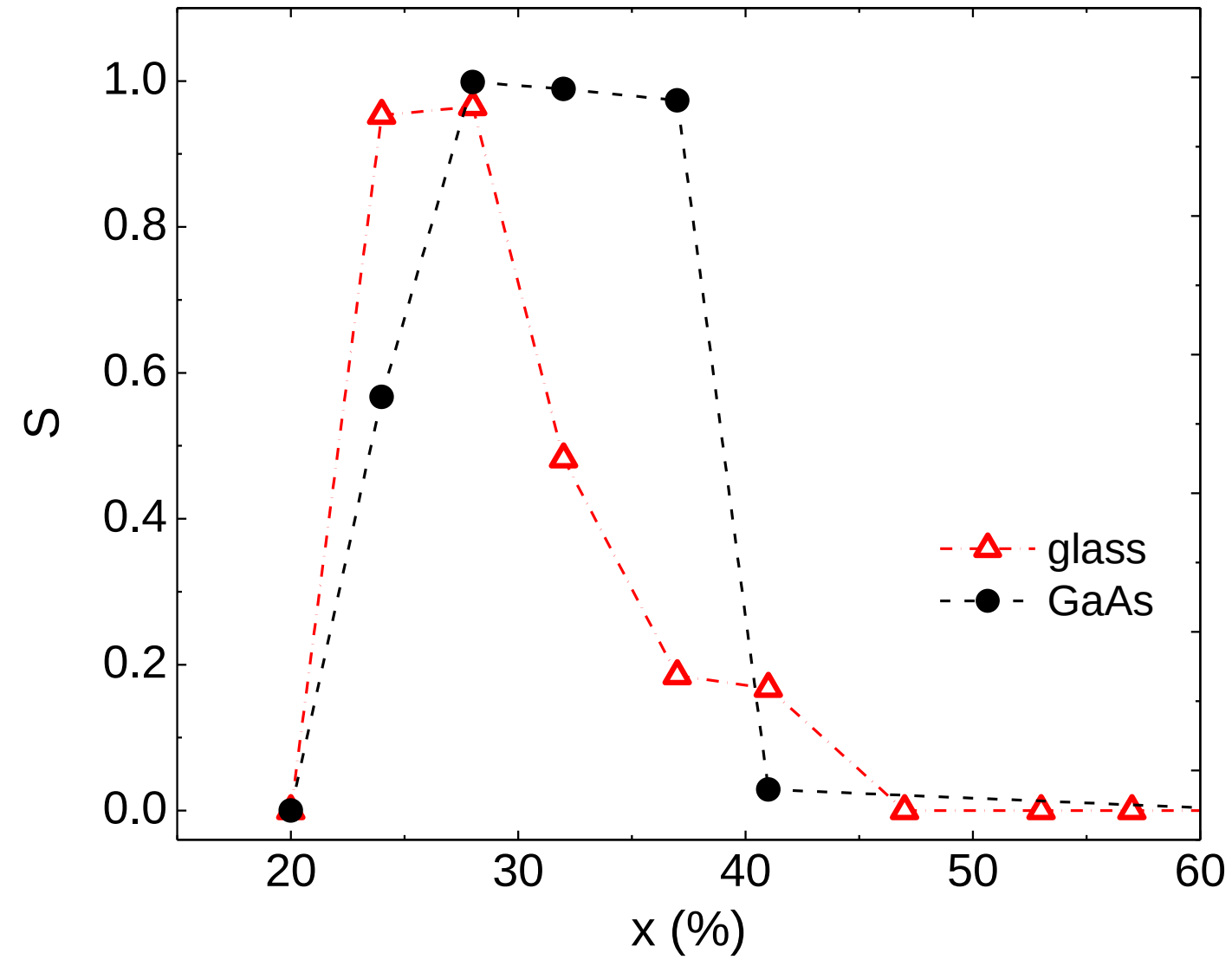

Fig. 4

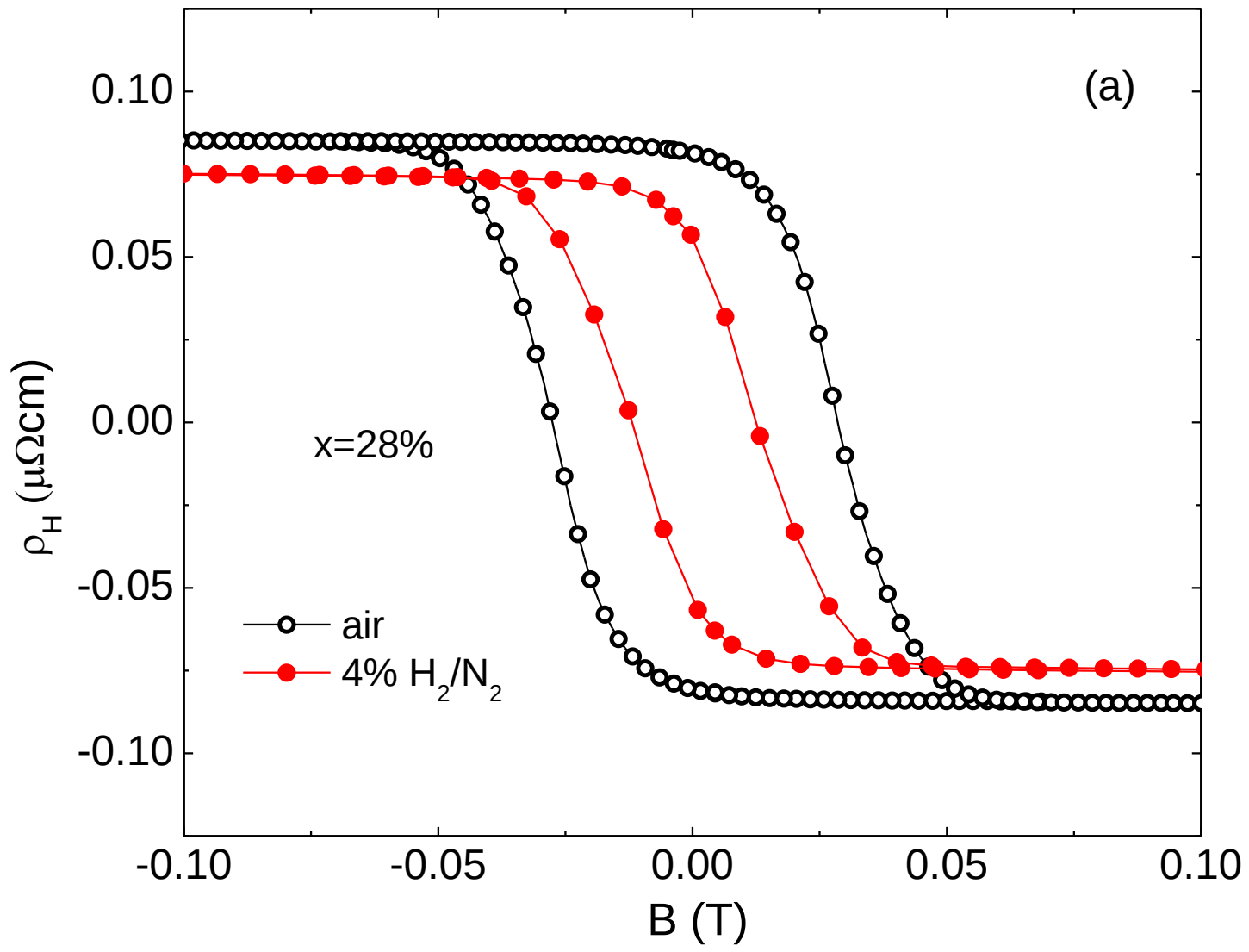

Fig. 5a

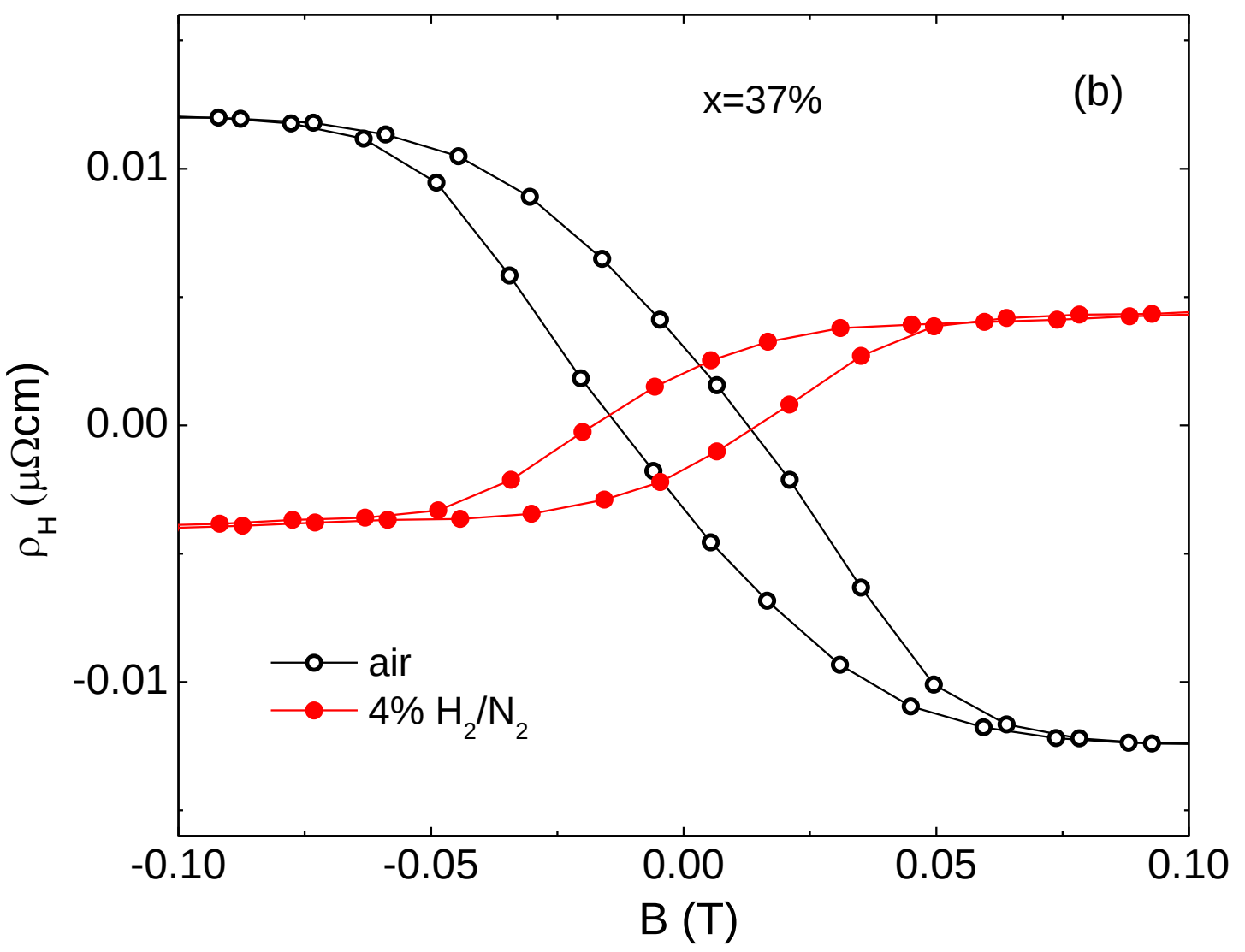

Fig. 5b

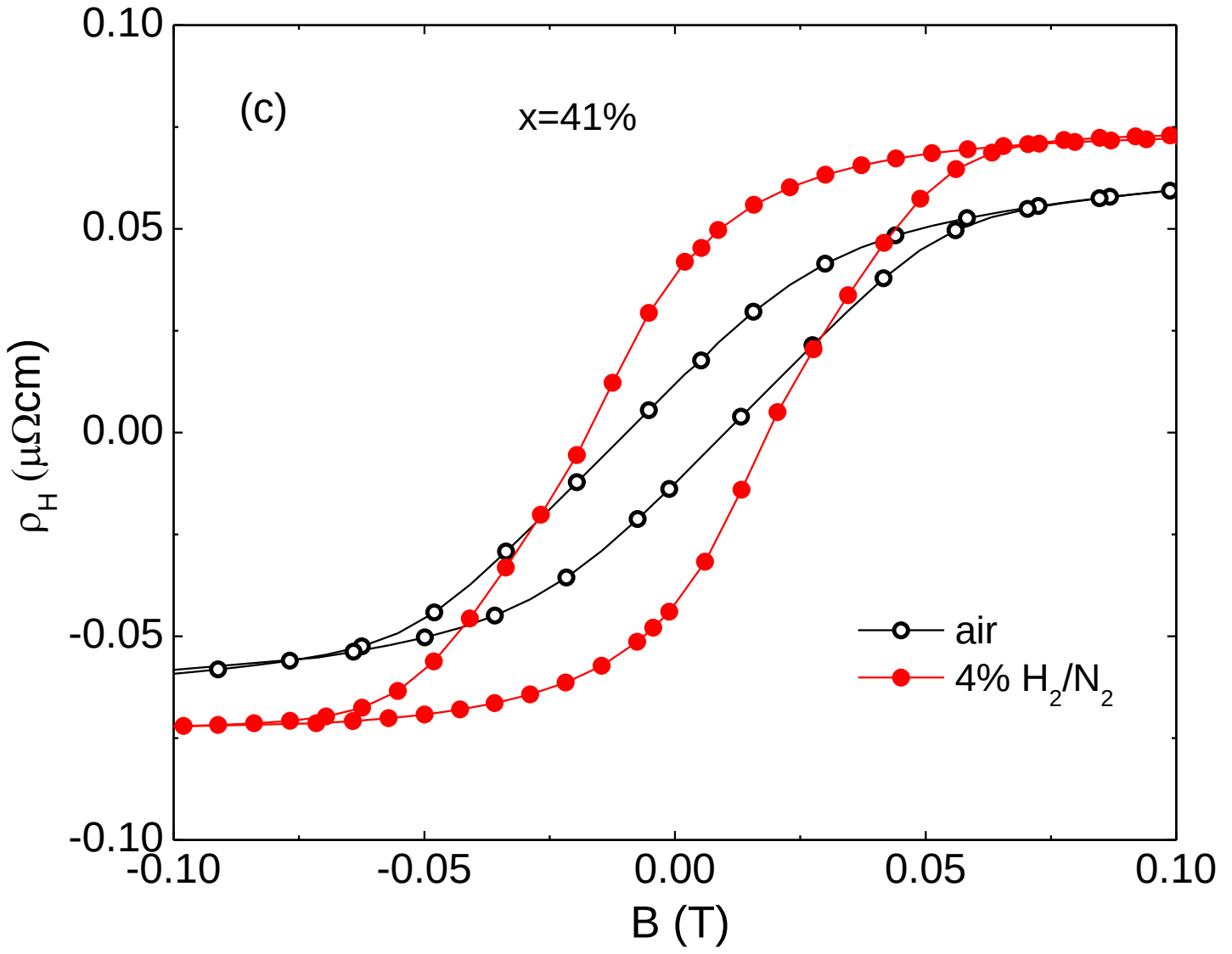

Fig. 5c

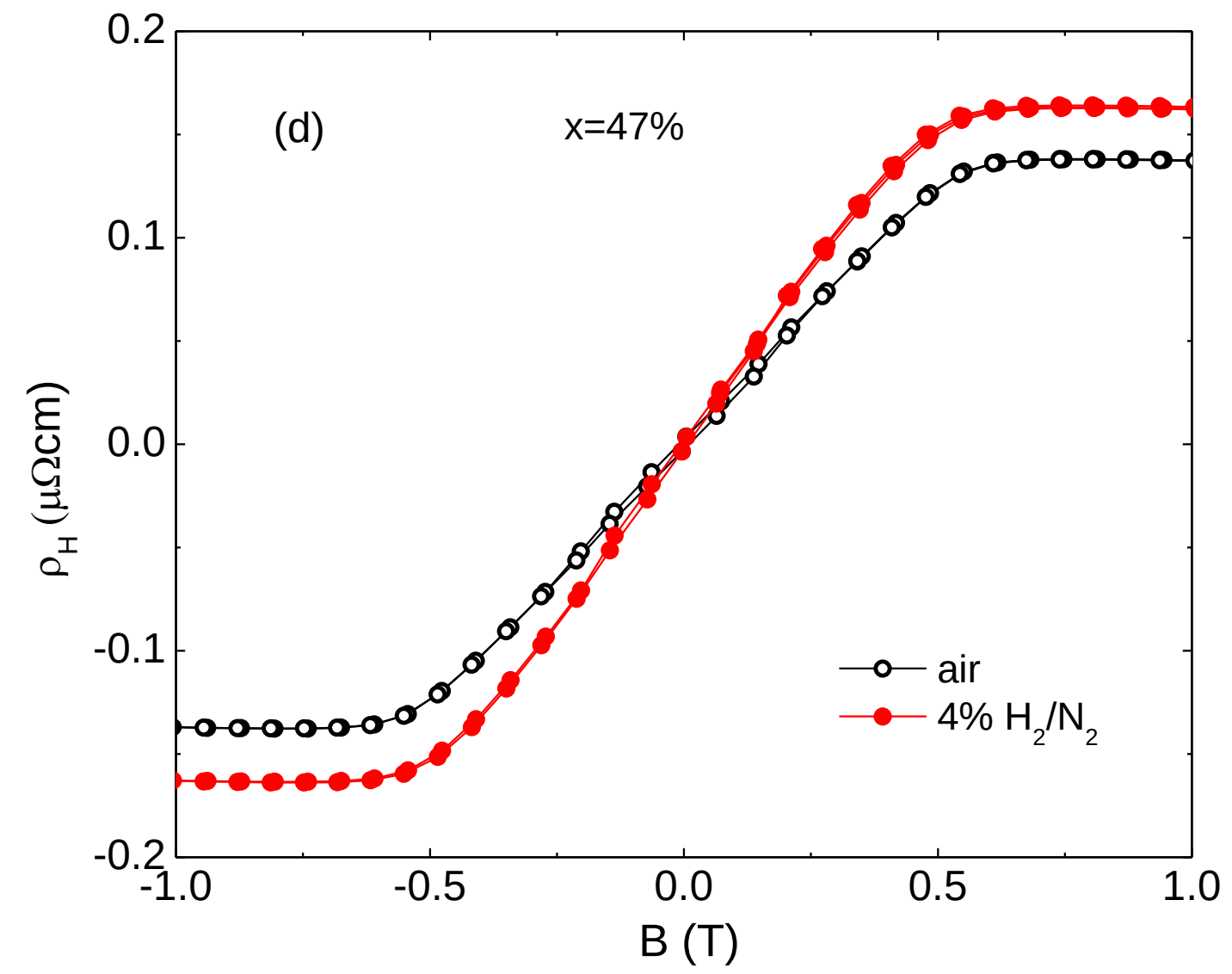

Fig. 5d

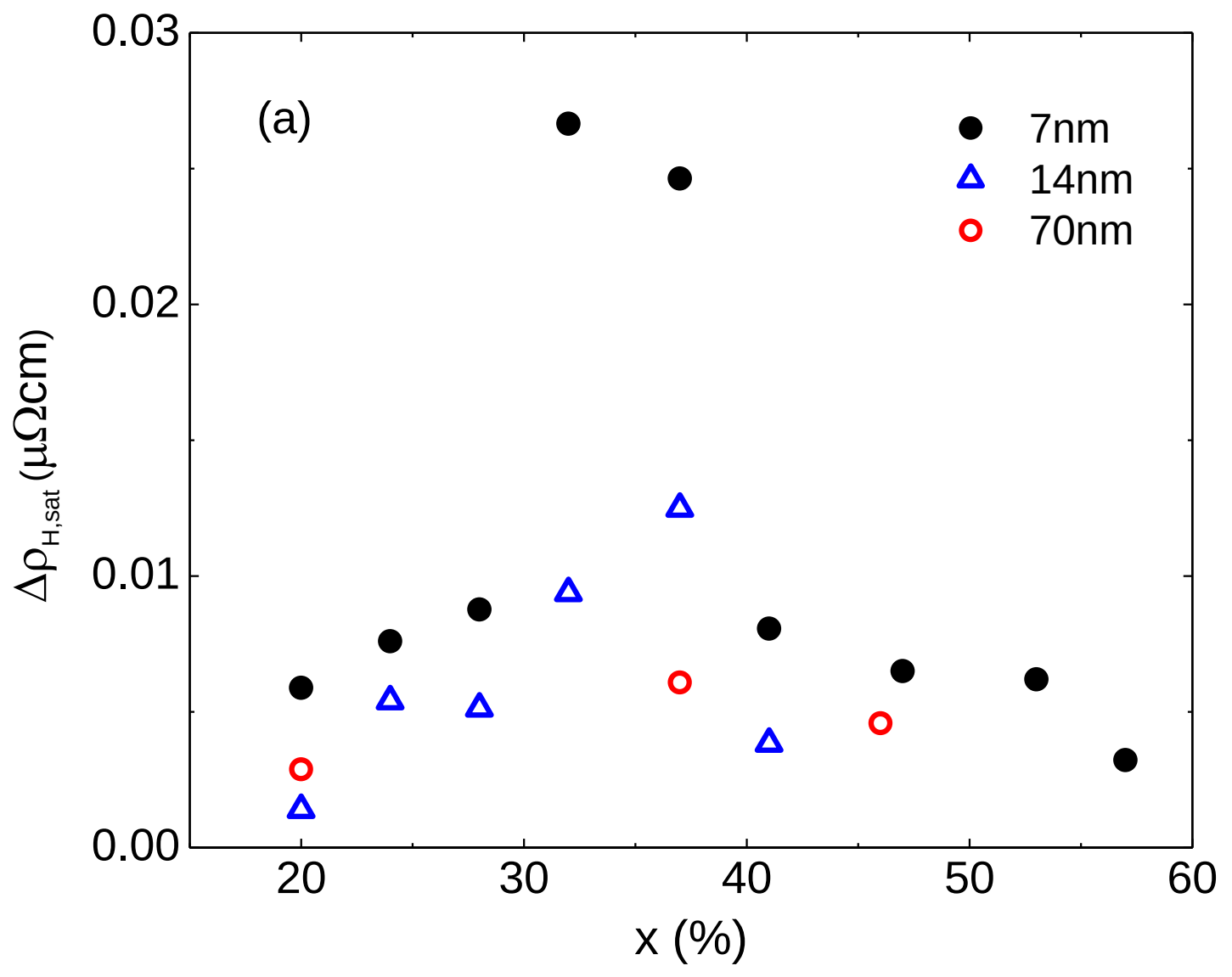

Fig. 6a

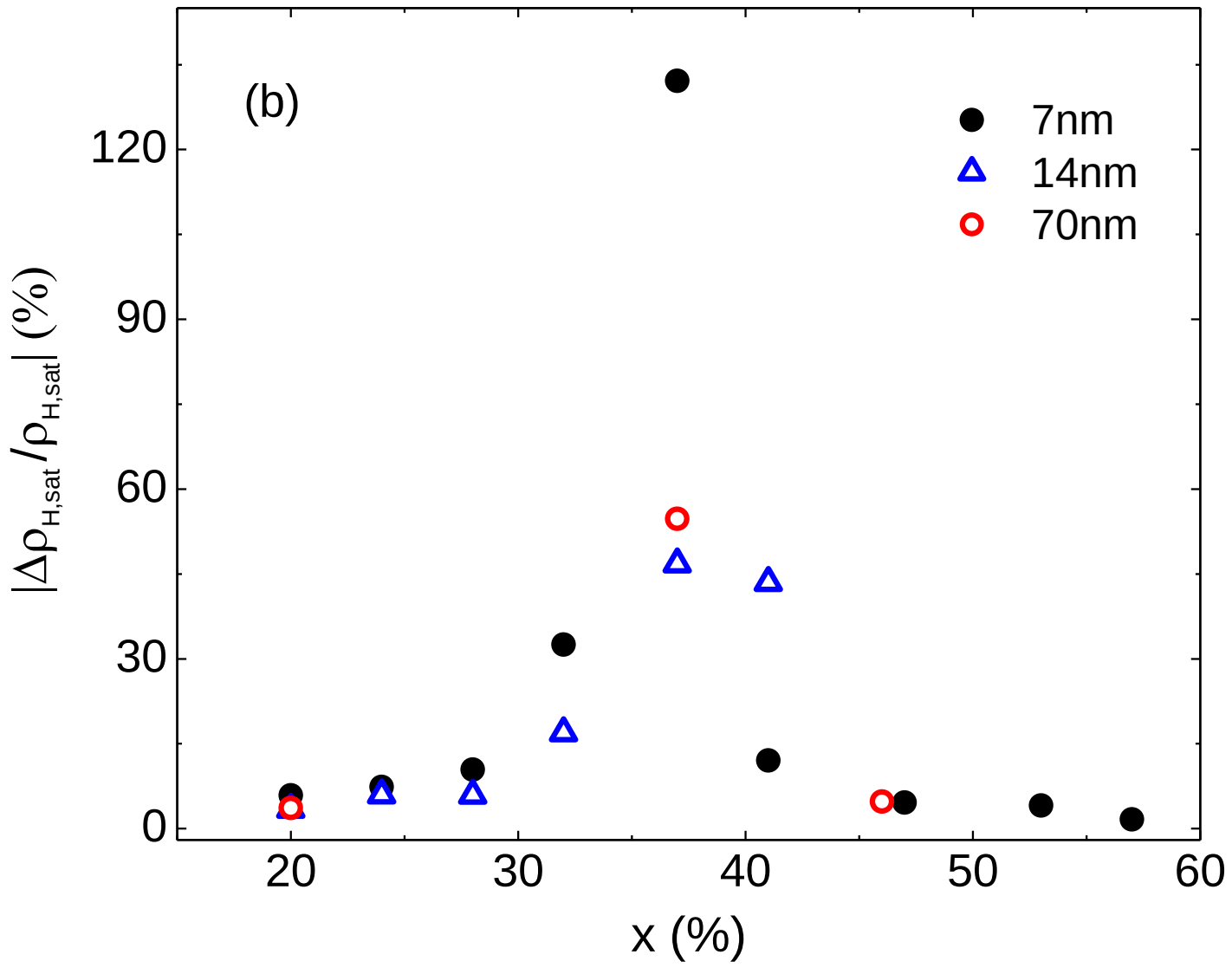

Fig. 6b

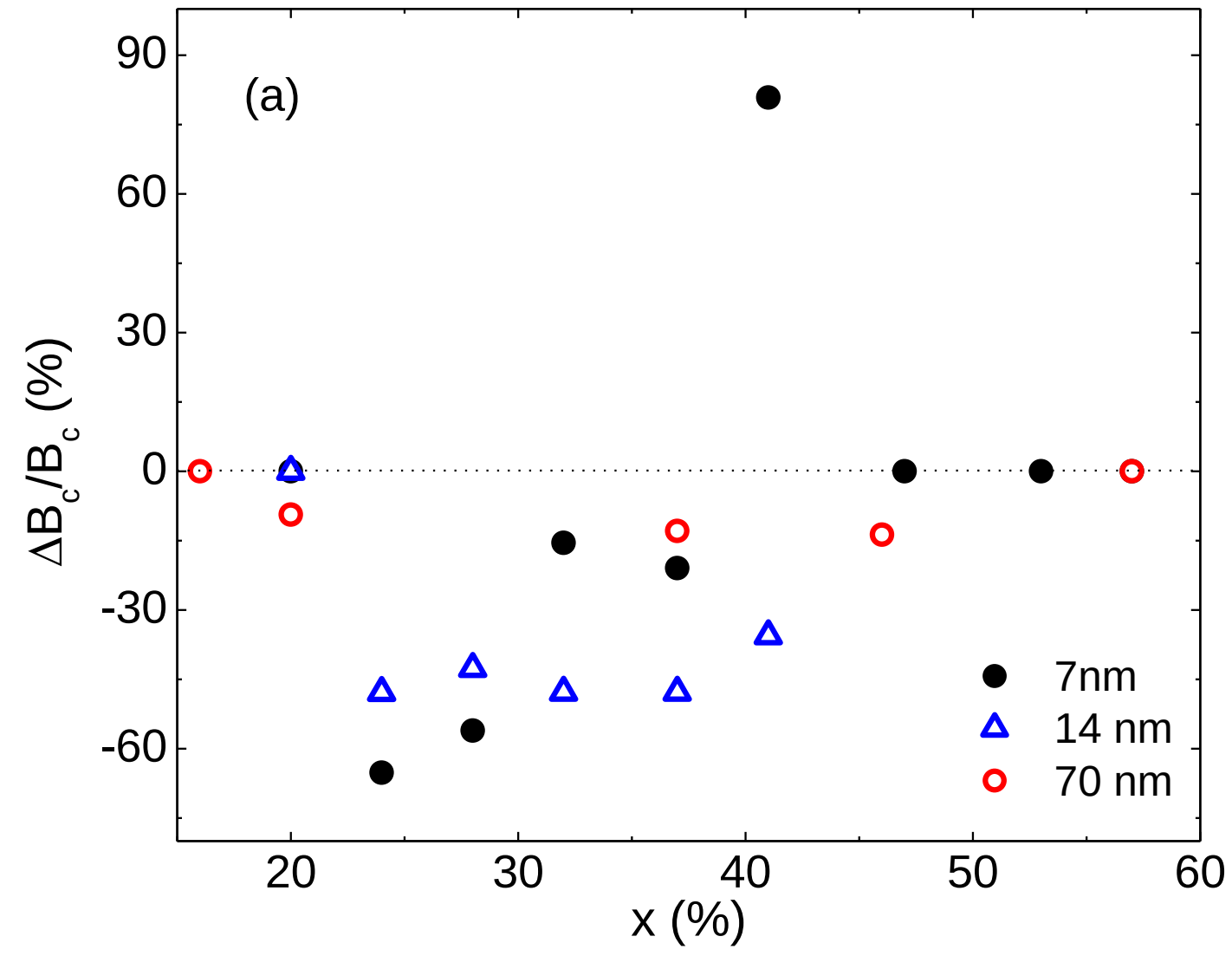

Fig. 7a

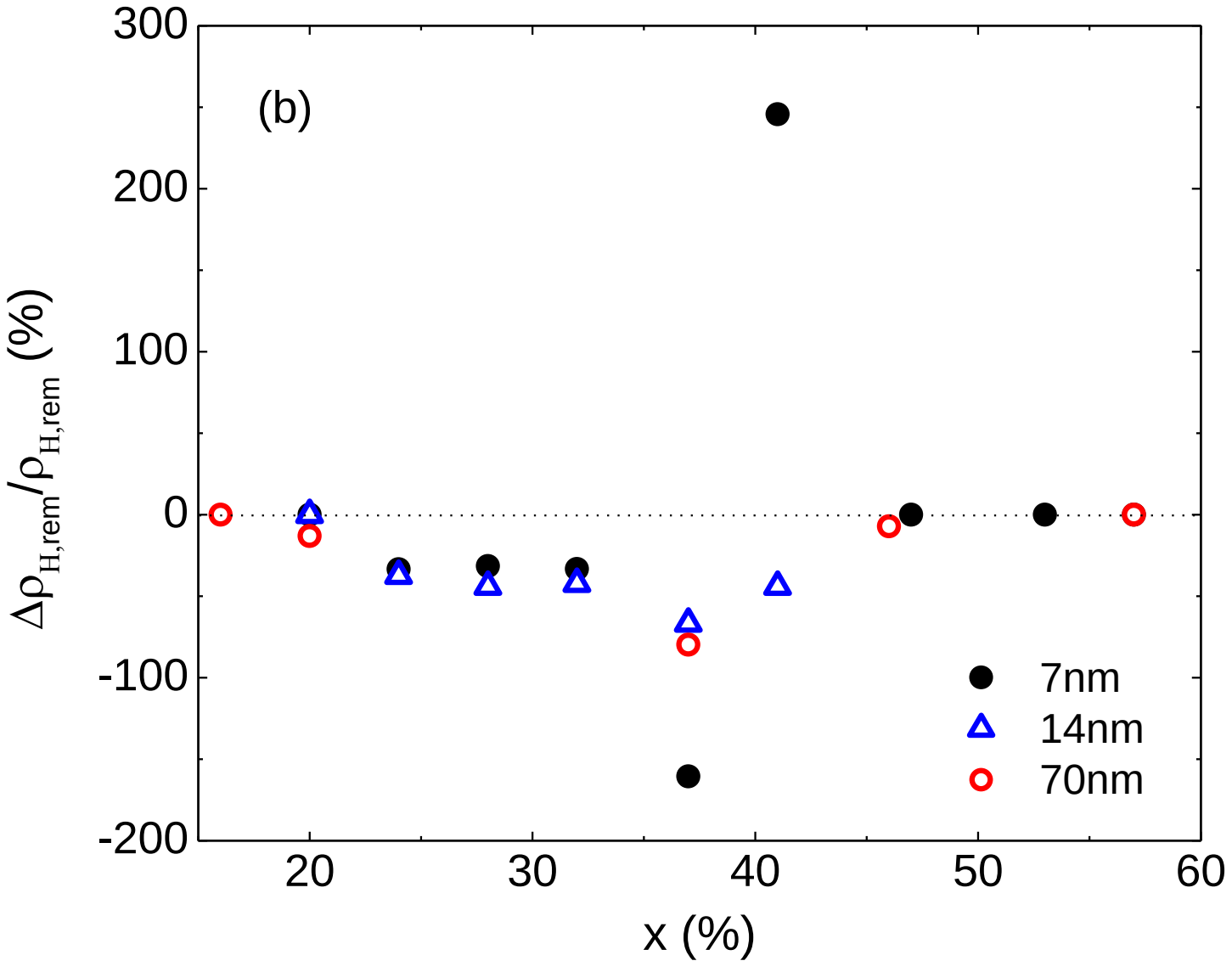

Fig. 7b

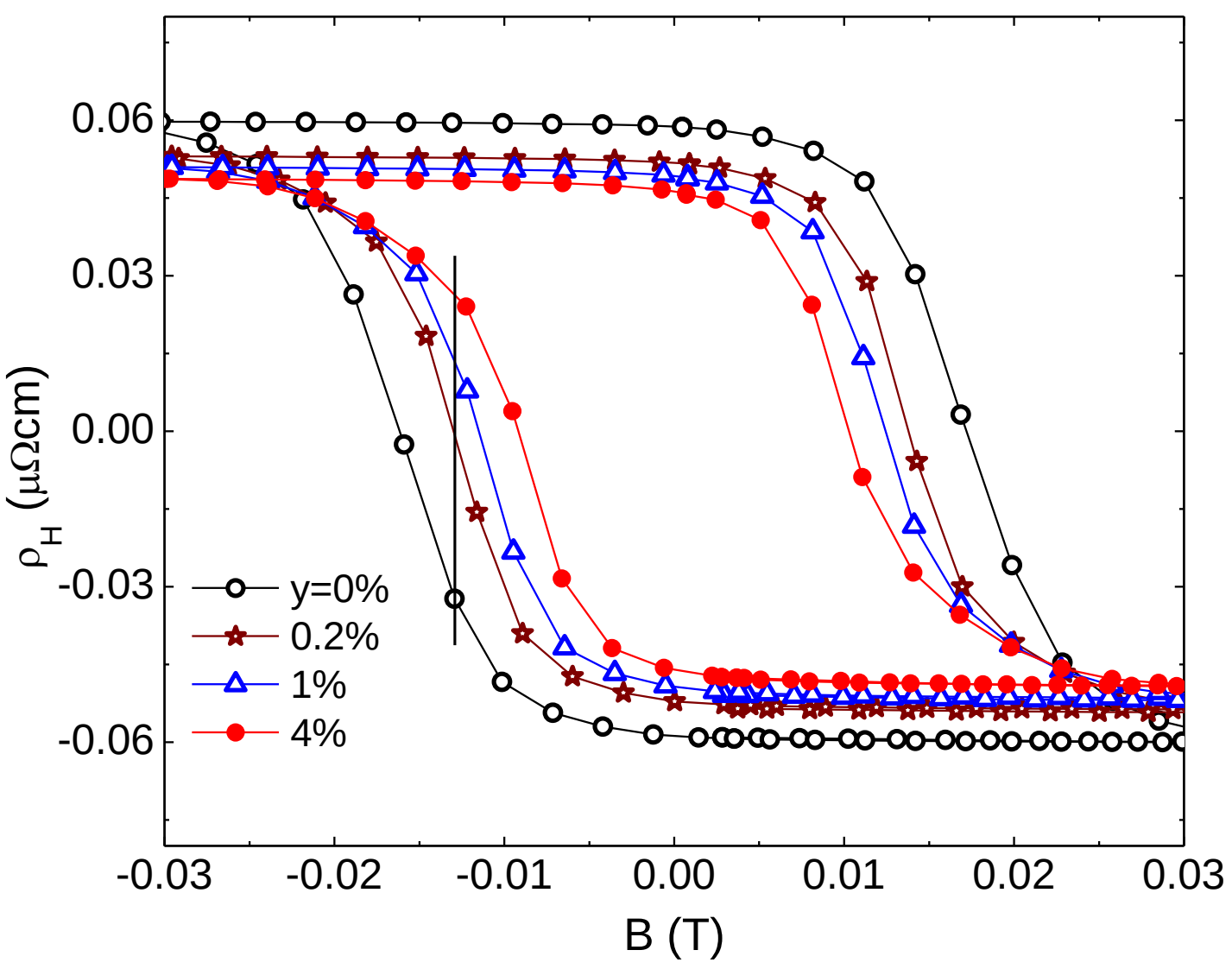

Fig. 8.

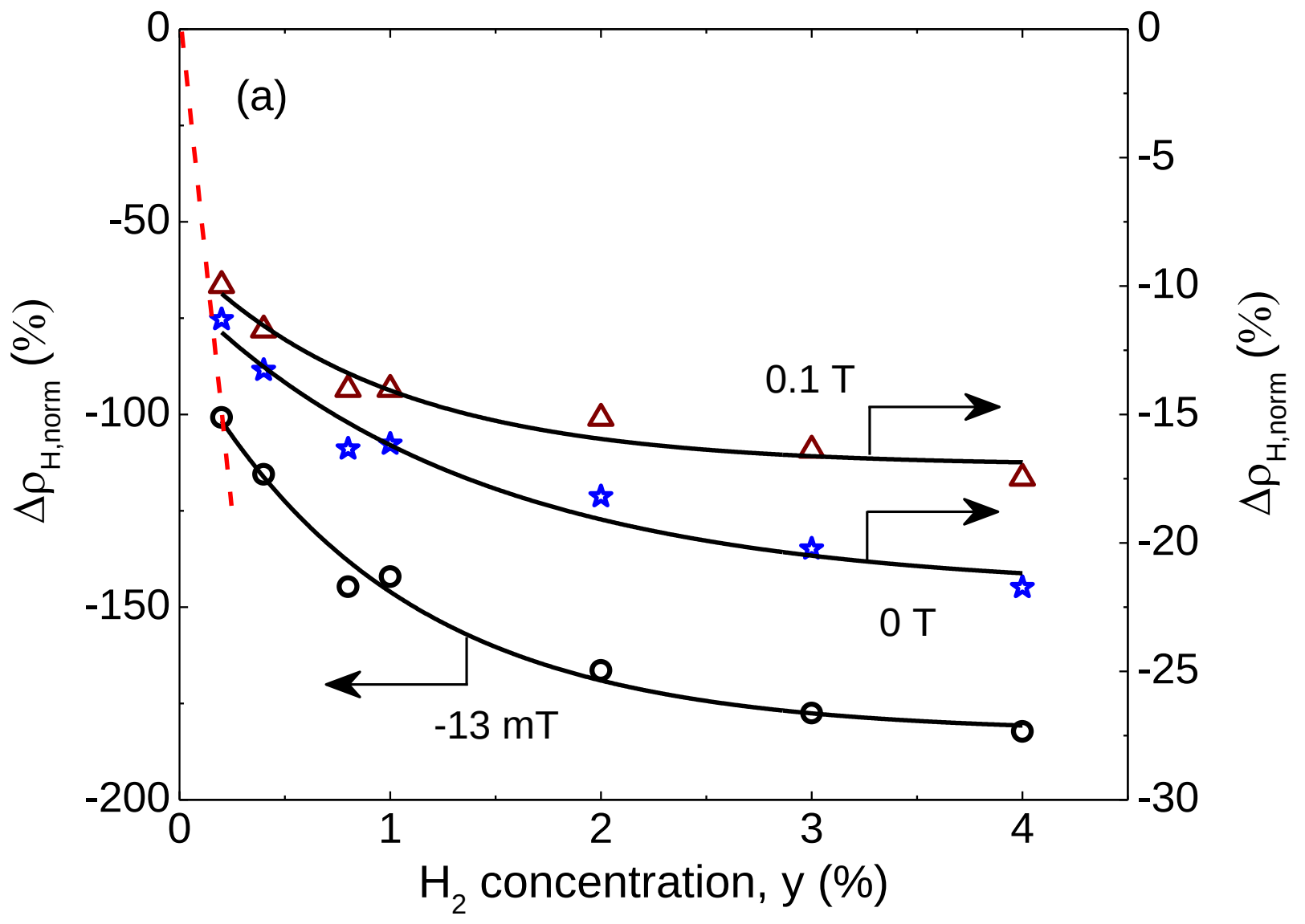

Fig. 9a

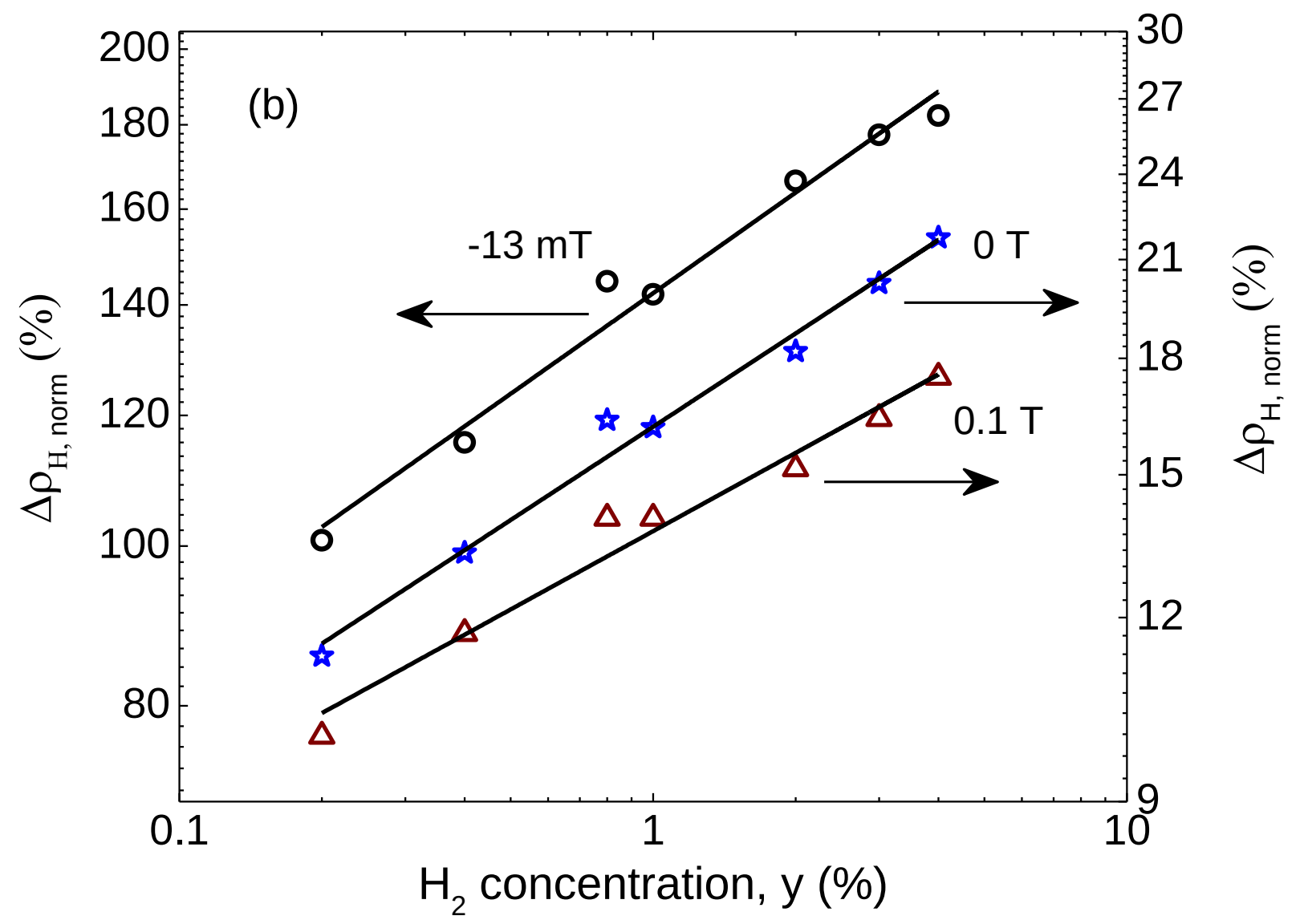

Fig. 9b

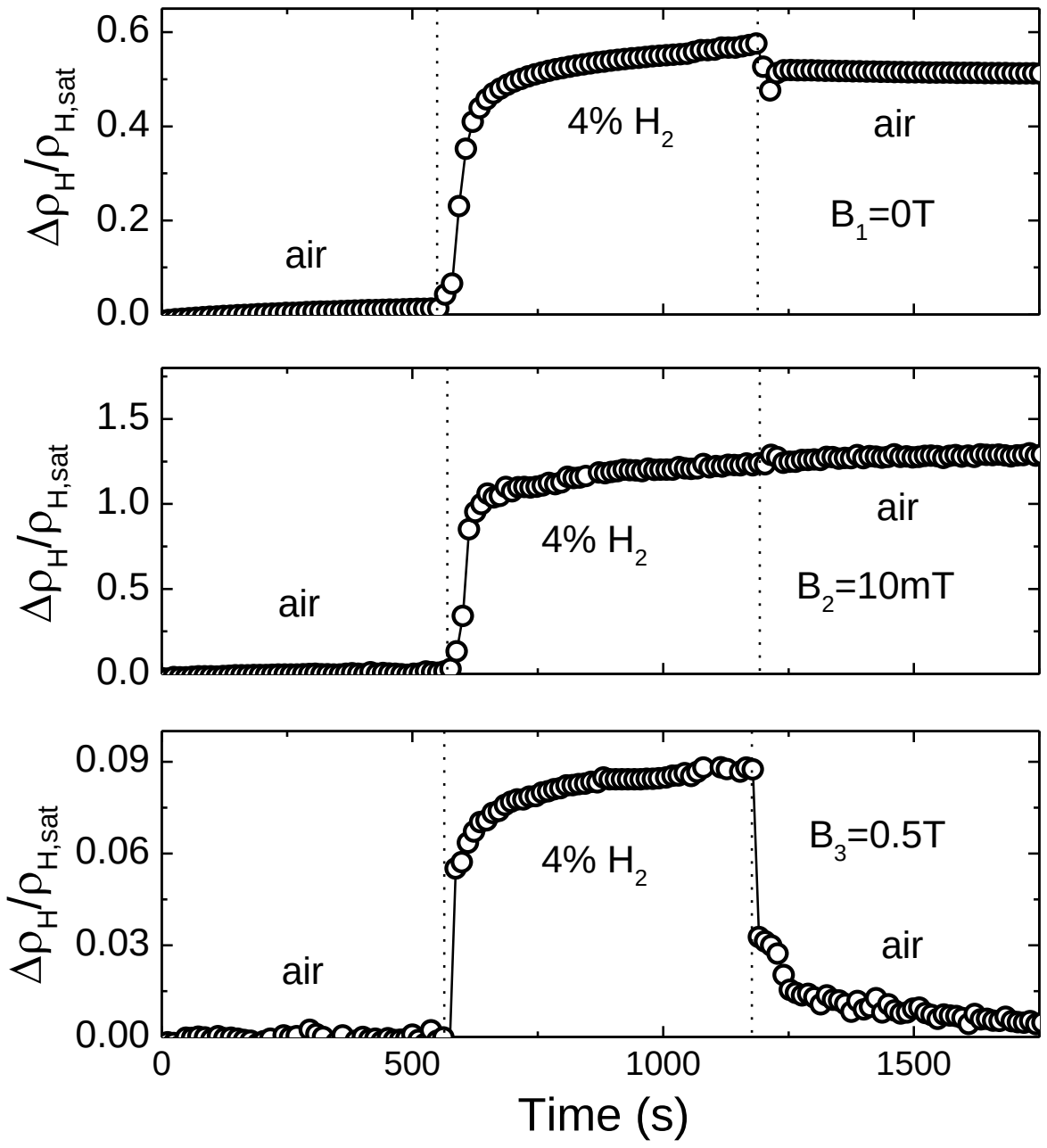

Fig. 10

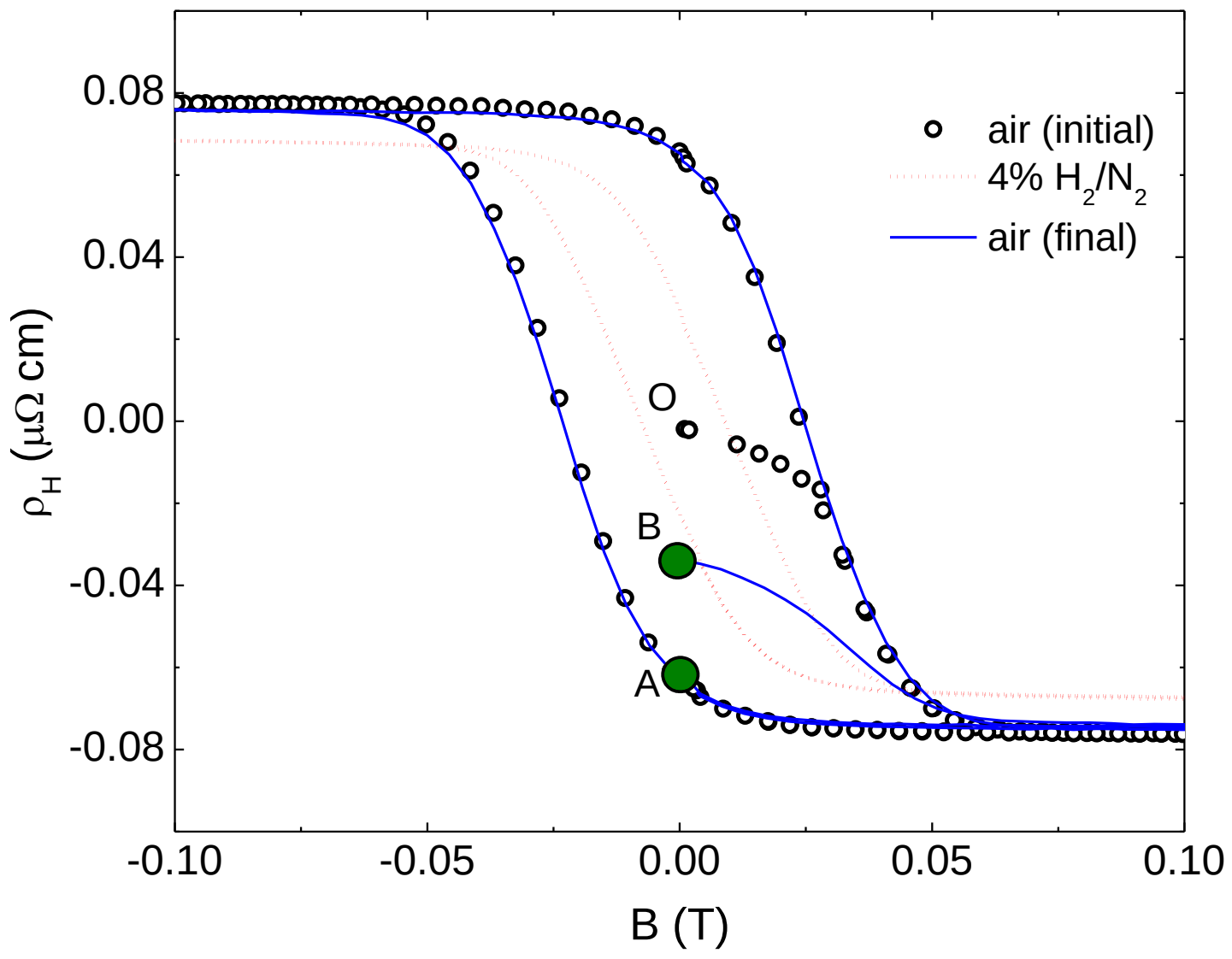

Fig. 11